\documentclass[letter]{aa}  
\usepackage{graphicx}
\usepackage{txfonts}
\usepackage[breaklinks, colorlinks, citecolor=blue, linkcolor=blue]{hyperref}
\usepackage{xspace}
\usepackage{xcolor}
\usepackage{booktabs}
\usepackage{lineno}

\newcommand{\nstar}{KELT-1\xspace}
\newcommand{\nplanet}{KELT-1b\xspace}

\newcommand{\pytransit}{\textsc{PyTransit}\xspace}

\newcommand{\celerite}{\textsc{Celerite}\xspace}

\newcommand{\tess}{\textrm{TESS}\xspace}
\newcommand{\spitzer}{\textit{Spitzer}\xspace}

\newcommand{\cheops}{\textrm{CHEOPS}\xspace}

\newcommand{\np}[2]{\ensuremath{N(#1,\,#2)}}

\newcommand{\rev}[1]{#1} 

\begin{document} 
   \title{Temporal albedo variability in the phase curve of \nplanet}

   \author{H. Parviainen\inst{1,2}\fnmsep\thanks{hannu@iac.es}}
   \institute{Instituto de Astrof\'isica de Canarias (IAC), E-38200 La Laguna, Tenerife, Spain
         \and Departamento de Astrof\'isica, Universidad de La Laguna (ULL), E-38206 La Laguna, Tenerife, Spain}

   \date{Received September 15, 1996; accepted March 16, 1997}
 
\abstract{
The dayside brightness spectrum of a highly irradiated transiting brown dwarf \nplanet has been shown to be challenging to explain with the current brown dwarf atmosphere models. The spectrum has been measured from observations spanning ten years and covering high-precision secondary eclipses and phase curves from space in blue-visible (CHaracterising ExOPlanet Satellite, \cheops), red-visible (Transiting Exoplanet Survey Satellite, \tess), and near-infrared (\spitzer), as well as secondary eclipse observations in near-infrared from the ground. First, the dayside of \nplanet was observed to be brighter in the \tess passband than expected based on earlier near-infrared phase curve observations with \spitzer, and recently, the dayside was observed to be extremely dark in the \cheops passband. While several theories have been proposed to reconcile the discrepancy between the \tess and \spitzer bands, explaining the difference between the largely overlapping \cheops and \tess bands has proven to be more difficult. 
   
Here, I model the \tess photometry from Sector 17 together with the new \tess photometry from Sector 57 and show that the discrepancies in \nplanet's dayside brightness spectrum are best explained by temporal variability in \nplanet's albedo. This variability is most likely due to changing silicate cloud coverage on the brown dwarf's dayside, that is, weather.
}


\maketitle

\section{Introduction} 
\label{sec:introduction}

\nplanet \citep{Siverd2012} is a strongly irradiated brown dwarf transiting a 6500~K F5-star on a short-period orbit of 1.2~days. The brown dwarf's dayside brightness spectrum has been measured using secondary eclipse and phase curve observations from space in blue-visible \citep[CHaracterising ExOPlanet Satellite, \cheops,][]{Parviainen2022a}, red-visible \citep[Transiting Exoplanet Survey Satellite, \tess,][]{Beatty2020, VonEssen2021}, and near-infrared \citep[\spitzer,][]{Beatty2014, Beatty2019a}, as well as in near-infrared (NIR) from the ground \citep{Croll2015, Beatty2017}. 

Originally, \nplanet's dayside brightness temperature estimates from the ground-based eclipse observations by \citet{Croll2015} and \citet{Beatty2017} agreed reasonably well with the estimate of $\sim$2900~K from the \spitzer eclipse and phase curve observations in NIR\footnote{The original dayside brightness estimates from the secondary eclipse observations by \citet{Beatty2014}, \citet{Croll2015}, and \citet{Beatty2017} were biased due to a strong stellar ellipsoidal variation signal that was not included into the eclipse models, but these biases were taken into account in the later studies.} \citep{Beatty2014, Beatty2019a}. However, high-precision phase curve observations in red-visible by \tess (Sector 17) showed unexpectedly high dayside brightness that did not agree with the original atmosphere models based on \spitzer observations \citep{Beatty2020, VonEssen2021}. 

Two theories were proposed to reconcile the discrepancy between the \tess and \spitzer dayside brightnesses. First, \citet{Beatty2020} proposed the discrepancy to be due to a high albedo in the visible caused by silicate clouds. \nplanet's dayside is too hot for clouds to form, but its nightside temperature should be cool enough for the formation of silicate clouds that could be blown over from the nightside to the dayside by winds \rev{\citep[see][and the references therein]{Gao2021}}. Second, \citealt{VonEssen2021} proposed that the discrepancy can be explained by collision-induced absorption due to H$_2$-H$_2$ and H$_2$-He decreasing the brightness temperature in the \spitzer bands. That is, rather than the brightness of the \tess passband being boosted by reflection from clouds, the brightness in the \spitzer band would be suppressed by molecular absorption.

Recently, another discrepancy was observed between new blue-visible eclipse observations by \cheops and the red-visible observations by \tess. The \cheops observations suggested that \nplanet's dayside is very dark in the \rev{\cheops band, while the \tess observations showed it to be bright, even though the two passbands have a significant overlap \citep{Parviainen2022a}}. 

\rev{In this letter, I present a likely explanation for the abovementioned discrepancies based on a phase curve analysis of \nplanet using a \tess light curve from Sector 17 and a new \tess light curve from Sector 57. 
The data and code for the analysis and figures are publicly available from GitHub.\!\footnote{\url{https://github.com/hpparvi/2023_KELT-1b}.}}

\section{Observations}
\label{sec:observations}
The Transiting Exoplanet Survey Satellite \citep[\tess,][]{Ricker2014} observed \nplanet during Sector 17 for 25 days (2019 Oct 08--2019 Nov 02) and Sector 57 for 29 days (2022 Sep 9--2022 Oct 10) with a two-minute cadence.\!\footnote{Sector 57 would also have short-cadence photometry available, but a two-minute cadence is sufficient for the analysis presented here.} I chose to use the Presearch Data Conditioning (PDC-SAP) light curve \citep{Stumpe2014, Stumpe2012, Smith2012a} produced by the SPOC pipeline \citep{Jenkins2016}, and remove all the exposures with a non-zero quality flag. Otherwise, the PDC-SAP photometry is used in the analysis as-is without additional detrending. Instead, I use a \celerite-based \citep{Foreman-Mackey2017} time-dependent Gaussian process (GP) with loosely constrained hyperparameters to account for time-dependent variability in the analysis, as detailed later.

\section{Theory and Numerical Methods}
\label{sec:theory}

\rev{The posteriors for all physical quantities of interest are estimated using a standard Bayesian parameter estimation approach \citep{Parviainen2018} as in \citet{Parviainen2022a}.}
I model the \tess photometry for both Sectors simultaneously using a \pytransit-based phase curve model described in detail in \citet{Parviainen2022a}. The orbital parameters (zero epoch, impact parameter, stellar density, $\sqrt{e}\cos\omega$, and $\sqrt{e}\sin\omega$), Doppler beaming amplitude, ellipsoidal variation amplitude, and the brown-dwarf to star area ratio do not depend on the Sector, but the limb darkening coefficients,\!\footnote{Stellar limb darkening was allowed to differ between the two sectors to account for any possible limb darkening differences due to unocculted spots.} \rev{and phase curve parameters (maximum and minimum surface brightness ratios and hotspot offset) are allowed to be different for each Sector. The orbital parameters and the area ratio are given loosely constraining priors based on \citet{Parviainen2022a}.}

\rev{The GP uses the \texttt{Matern32Term} \celerite term that approximates the Matern-3/2 covariance kernel. The GP hyperparameters are Sector-dependent and constrained with loosely informative normal priors.}

\section{Results and discussion}
\label{sec:results}

\begin{figure}
    \centering
    \includegraphics[width=\columnwidth]{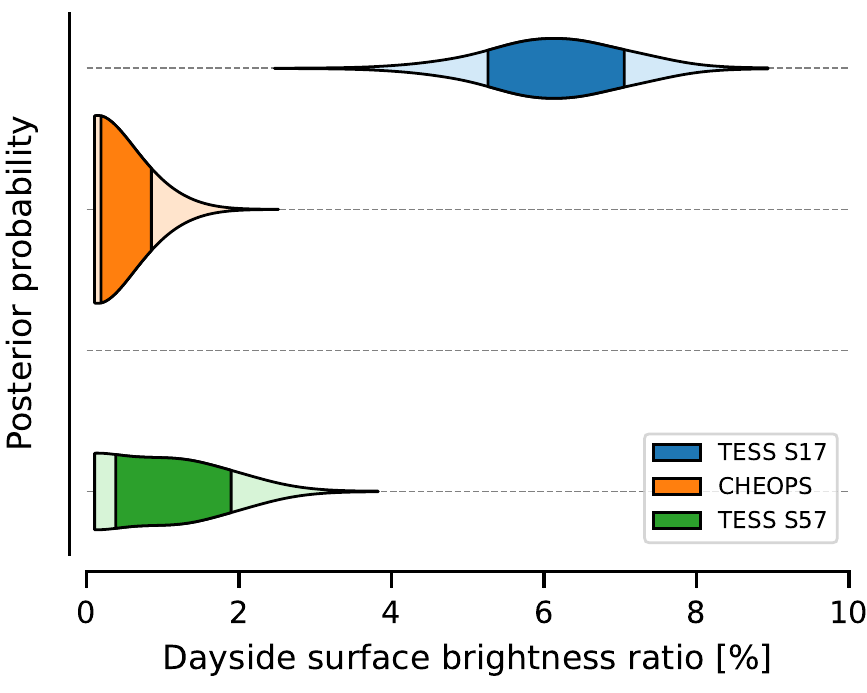}
    \caption{\rev{Dayside brown dwarf to star flux ratio} posteriors measured from \tess Sector 17 photometry, \cheops secondary eclipse observations, and \tess Sector 57 photometry. The y-axis offsets represent the average time differences between the observation sets (one year between the \tess S. 17 and \cheops observations and two years between the \cheops and \tess S. 57 observations).}
    \label{fig:brightness_ratios}
\end{figure}

\begin{figure*}
    \centering
    \includegraphics[width=\textwidth]{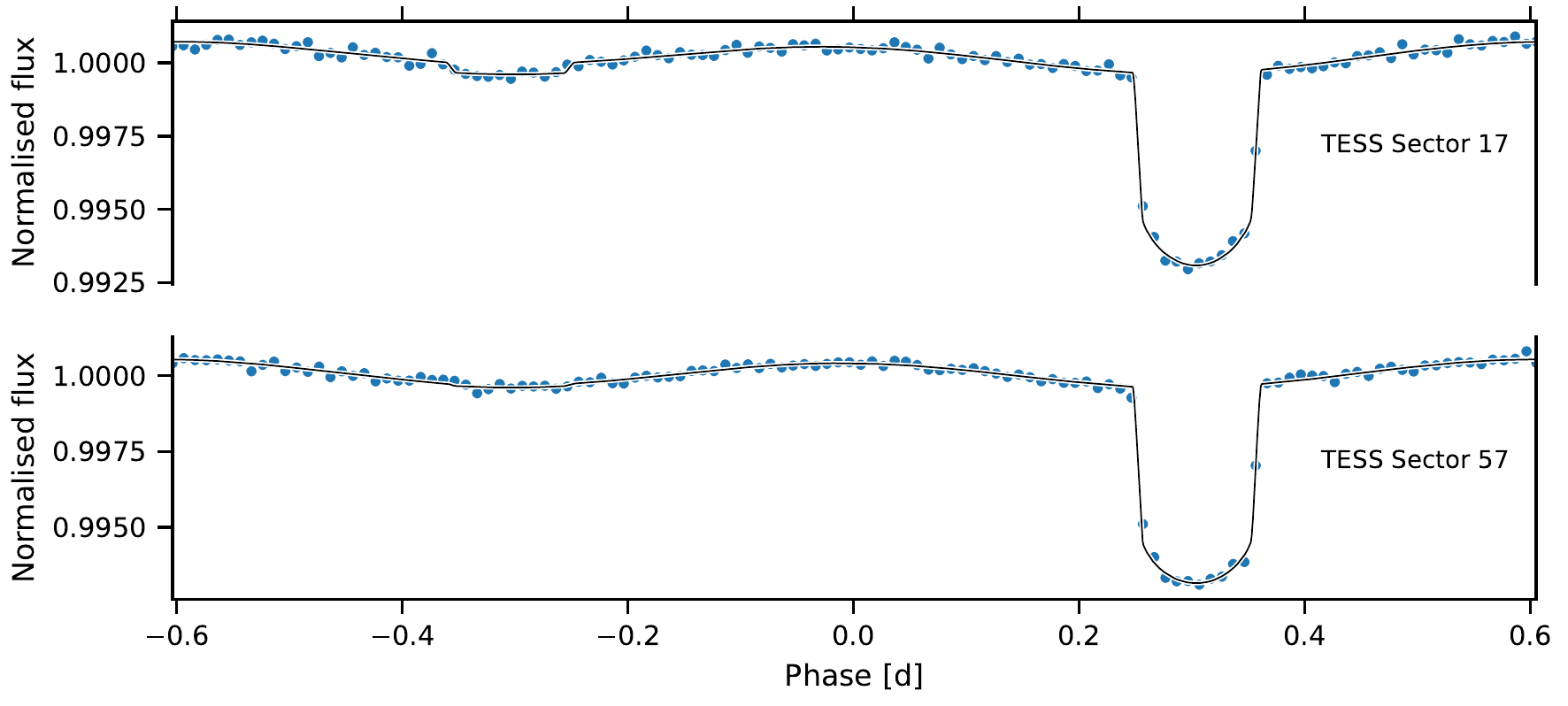}
    \caption{\tess photometry from Sectors 17 and 57 with the posterior GP variability model removed, folded over the \nplanet orbital phase and binned over 15 minutes as blue points (the errorbars are smaller than the points) and the median posterior phase curve model (with uncertainties smaller than the line width).}
    \label{fig:phase_curve}
\end{figure*}

\begin{figure*}
    \centering
    \includegraphics[width=\textwidth]{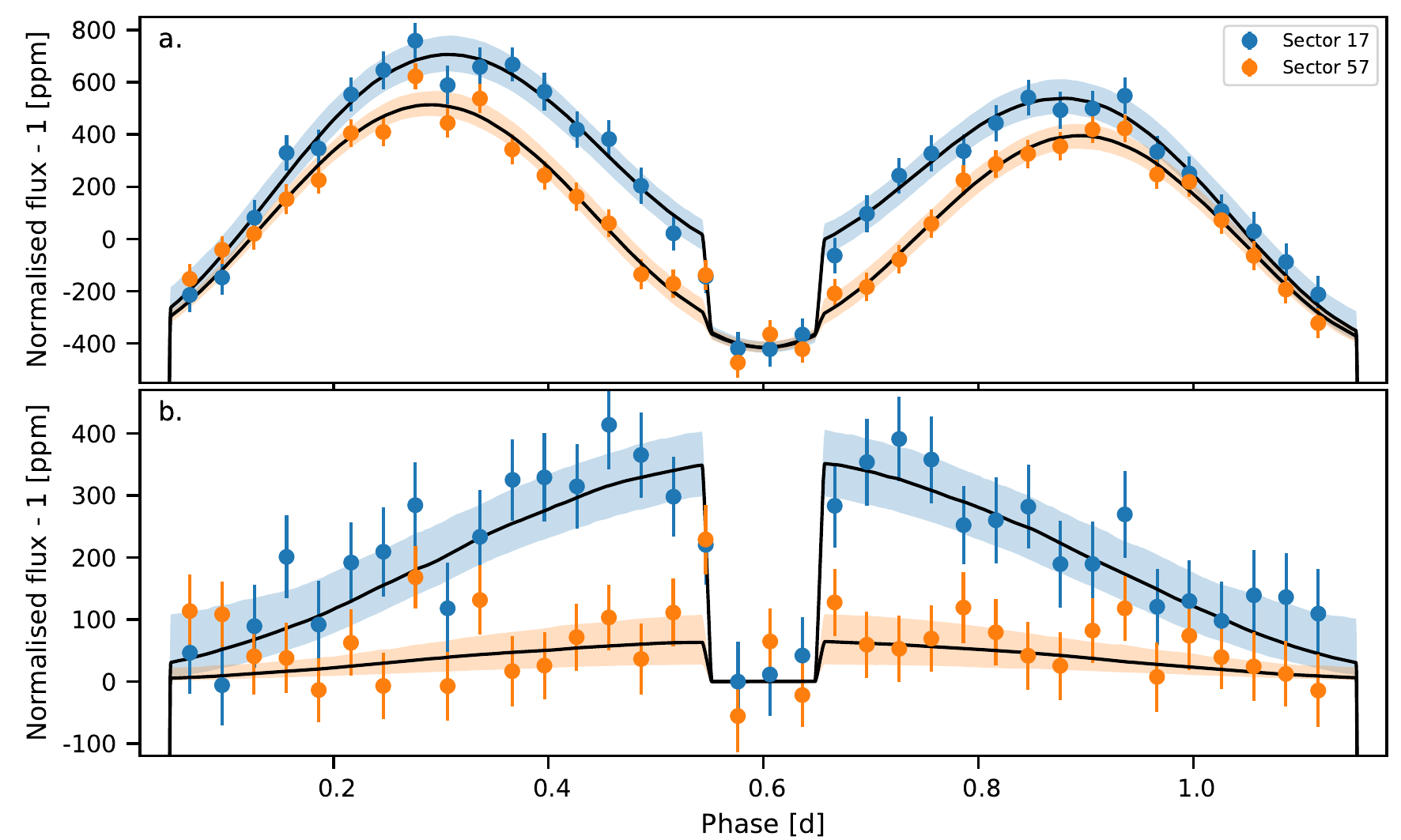}
    \caption{As in Fig.~\ref{fig:phase_curve}, but centred around the eclipse and binned over 43 minutes (panel a.), and with the stellar ellipsoidal variation and Doppler beaming signals removed to show only the phase curve signal from \nplanet (panel b).}
    \label{fig:zoomed_phase_curve}
\end{figure*}

I list the \rev{flux ratio}\footnote{\rev{As in \citet{Parviainen2022a}, the \emph{flux ratio} stands for the ratio of the radiation emitted towards the observer by the brown dwarf and the star per projected unit area. The \emph{flux ratio} is not the ratio of the disk-integrated fluxes, but this can be obtained by multiplying the flux ratio by the companion-star area ratio, $k^2$.}} posterior estimates in Table~\ref{table:flux_ratio_posteriors}, \rev{the GP hyperparameter priors and posteriors in Table~\ref{table:priors_and_posteriors}}, show the surface brightness ratio posteriors in Fig.~\ref{fig:brightness_ratios} together with the \cheops posterior estimate from \citet{Parviainen2022a}, and show the posterior phase curve models in Figs.~\ref{fig:phase_curve}~and~\ref{fig:zoomed_phase_curve}. The area ratio, orbital parameters, ellipsoidal variation, and Doppler beaming estimates all agree with the previous results by \citet{Parviainen2022a} and are not listed.

\begin{table}
    \centering
    \caption{Dayside flux ratios and eclipse depths with their uncertainties.}
    \label{table:flux_ratio_posteriors}
    \begin{tabular*}{\columnwidth}{@{\extracolsep{\fill}} lccccc}
	\toprule\toprule
    Passband &  Dayside flux ratio [\%] & Eclipse depth [ppm] \\
	\midrule     
    \cheops\tablefootmark{a}     &        $< 1.50$ &          $< 90$ \\
    \tess Sector 17 &  $ 6.1 \pm 0.9$ &  $  360 \pm 50$ \\
    \tess Sector 57 &  $ 1.3 \pm 0.7$ &  $  74 \pm  42$ \\
    \bottomrule
    \end{tabular*}
    \tablefoot{
    \tablefoottext{a}{From \citet{Parviainen2022a}}.}
\end{table}

\begin{table}
    \centering
    \caption{Priors and posteriors for the \celerite Matern-3/2 term hyperparameters and the white noise standard deviation.}
    \label{table:priors_and_posteriors}
    \begin{tabular*}{\columnwidth}{@{\extracolsep{\fill}} llr}
	\toprule\toprule
     Parameter &  Prior & Posterior \\
	\midrule
    S$_{17} \ln \sigma$  &  \np{-6.1}{1.0} &  $  -7.8 \pm 0.1$ \\
    S$_{17} \ln \rho$ &  \np{1.0}{1.0} &  $  -1.5 \pm 0.2$ \\
    S$_{17}$ log$_{10} \;\sigma_\mathrm{wn}$ &  \np{-2.918}{0.025} &  $  -2.912 \pm 0.003$ \\
    S$_{57} \ln \sigma$ & \np{-6.1}{1.0} &  $  -7.8 \pm 0.1$ \\
    S$_{57} \ln \rho$ &  \np{1.0}{1.0} &  $  -1.1 \pm 0.2$ \\
    S$_{57}$ log$_{10} \;\sigma_\mathrm{wn}$  &  \np{-2.941}{0.025} &  $  -2.933 \pm 0.003$ \\
    \bottomrule
    \end{tabular*}
\end{table}

Figure~\ref{fig:zoomed_phase_curve} visualises the differences between the \tess Sector 17 and 57 phase curves. The \tess Sector 17 observations were carried out in October 2019, the \cheops observations in October and November 2020, and the \tess Sector 57 observations in September 2022. While the differences between the \tess Sector 17 and \cheops brightnesses could be accounted for by exotic atmosphere chemistry (see discussion in \citealt{Parviainen2022a}), the differences between the two \tess sectors can be only explained by temporal variability. The most likely source of temporal variability that can cause the discrepancies comes from variability in the cloud cover that changes \nplanet's dayside albedo.

\rev{The variability likely occurs over relatively long time scales.} \citet{VonEssen2021} already studied possible time variability in the \tess light curve from Sector 17 but concluded that the short-time-scale variability in phase curve shapes they discovered was due to stellar variability that disappeared after the stellar variability was taken into account \citep{VonEssen2021}. \cite{Parviainen2022a} repeated the eclipse-to-eclipse variability analysis using a GP to model the stellar variability and concluded that the per-eclipse SNR was too low for both \tess and \cheops to detect any short-term variability in the eclipse depths. However, the individual eclipse depths did not show any long-term trends in a time scale of two months covered by the \cheops observations.

\rev{Using a GP to model the time-dependent variability comes with a risk of the GP fitting the phase curve signal. This is especially the case since I allow the hyperparameters to differ for each Sector. I tested whether overfitting by GP could explain the difference between the two Sectors by carrying out a separate analysis without a GP. The analysis does not model the time-dependent variability in any way and is robust to overfitting. The test analysis results support the primary analysis,\!\footnote{\url{https://github.com/hpparvi/2023_KELT-1b/blob/main/A01b_white_noise_plots.ipynb}} and I conclude that the GP does not overfit the data in the primary analysis.}

\begin{figure*}
    \centering
    \includegraphics[width=\textwidth]{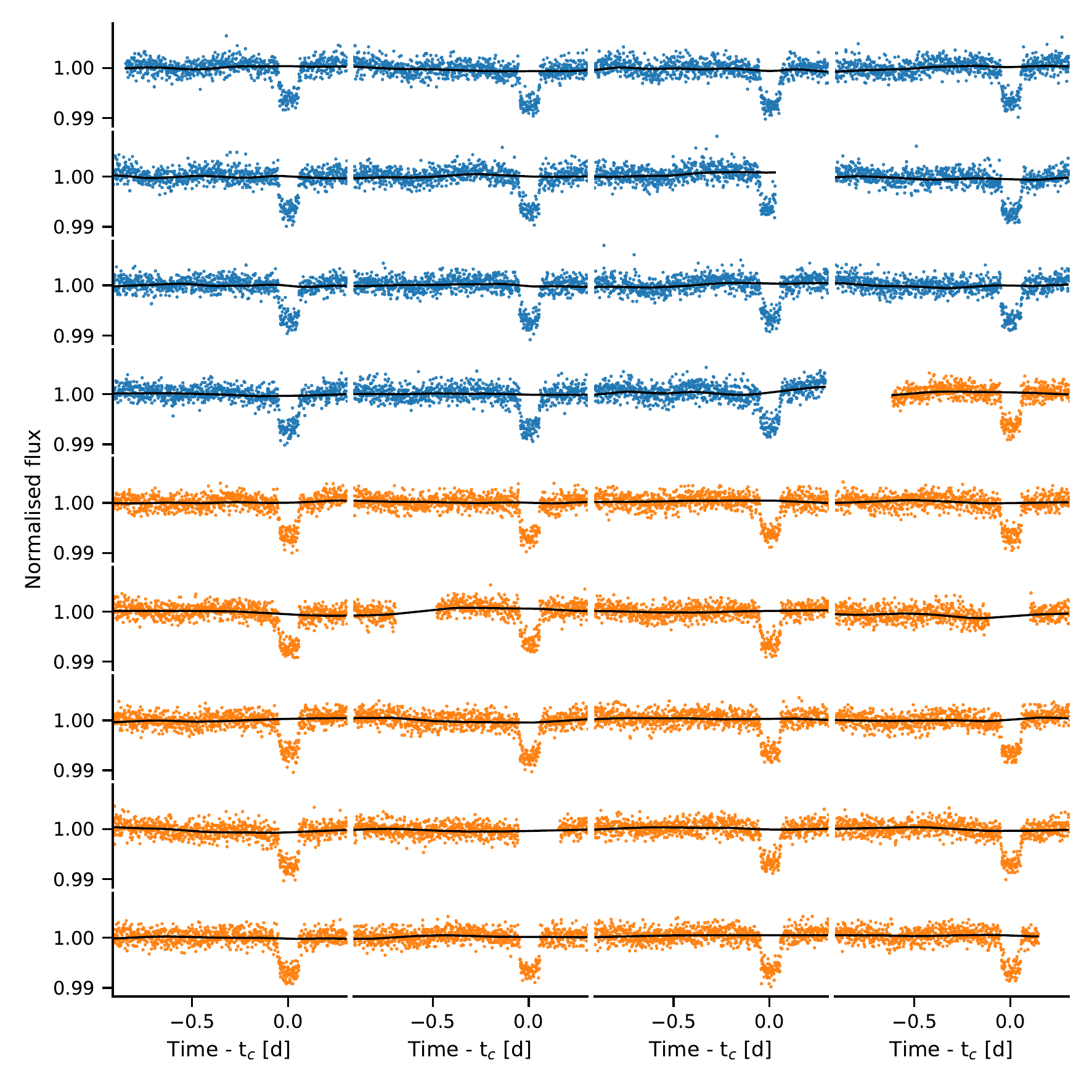}
    \caption{The TESS Sector 17 and 57 light curves (blue and yellow dots, respectively) with the GP baseline model drawn as a black line.}
    \label{fig:individual_light_curves}
\end{figure*}

\rev{As a sanity check, I repeated the analysis separately for each Sector to see whether the parameters assumed to be Sector-independent parameters showed any variability. All the orbital parameters, the area ratio, and ellipsoidal variation amplitude agreed within uncertainties between the two Sectors. Thus, Stellar variability is unlikely to be behind the discrepancy. Additionally, \citet{Siverd2012} find \nstar to be a relatively quiet star with rms scatter of 9.8 mmag measured over an observing period of 4.2 years, and the \tess light curves do not show any significant variability, as shown in Fig.~\ref{fig:individual_light_curves}.}

\section{Conclusions} 
\label{sec:conclusions}
I have modelled the \nplanet phase curves from \tess Sectors 17 and 57, separated by three years, and show that \nplanet's dayside brightness varies significantly between the two \tess Sectors. The difference can be best explained by temporal albedo variability caused by changing silicate cloud coverage on the brown dwarf's dayside since the two light curves were observed in the same passband. Weather is then the most likely solution also to explain the discrepancies between the \tess and \cheops observations reported by \citet{Parviainen2022a} and the \spitzer and \tess observations reported by \citet{Beatty2020} and \citet{VonEssen2021}. 

\begin{acknowledgements} 
\rev{I thank the anonymous referee for their timely and helpful comments and the \cheops community for the valuable discussion.
I acknowledge the support by the Spanish Ministry of Science and Innovation with the Ramon y Cajal fellowship number RYC2021-031798-I.}
I acknowledge the use of public \tess Alert data from pipelines at the \tess Science Office and at the \tess Science Processing Operations Center. 
Resources supporting this work were provided by the NASA High-End Computing (HEC) Program through the NASA Advanced Supercomputing (NAS) Division at Ames Research Center for the production of the SPOC data products.
\end{acknowledgements} 

\bibliography{kelt1}{}
\bibliographystyle{aa}




\end{document}